\documentclass[11pt]{article}
\usepackage{amsmath,amssymb,color,graphics,epsfig,cite}

\usepackage{amsfonts}

\newcommand{\be}{\begin{equation}}
\newcommand{\ee}{\end{equation}}
\newcommand{\bea}{\setlength\arraycolsep{2pt} \begin{eqnarray}}
\newcommand{\eea}{\end{eqnarray}}

\def\fft#1#2{{\frac{#1}{#2}}}

\def\0{{\sst{(0)}}}
\def\1{{\sst{(1)}}}
\def\2{{\sst{(2)}}}
\def\3{{\sst{(3)}}}
\def\4{{\sst{(4)}}}
\def\5{{\sst{(5)}}}
\def\6{{\sst{(6)}}}
\def\7{{\sst{(7)}}}
\def\8{{\sst{(8)}}}
\def\sst#1{{\scriptscriptstyle #1}}

\begin{document}

\begin{center}
{\Large {\bf Thermodynamics of Taub-NUT-AdS Spactimes}}

\vspace{20pt}

{\large Jun-Fei Liu and Hai-Shan Liu}

\vspace{10pt}

\vspace{10pt}

{\it Center for Joint Quantum Studies and Department of Physics,\\
School of Science, Tianjin University, Tianjin 300350, China }

\vspace{40pt}

\underline{ABSTRACT}

\end{center}

We apply the generalised Komar method proposed in [arxiv:2208.05494] to Taub-NUT-AdS spacetime in the theory of Einstein gravity plus a cosmological constant. Based on a generalised closed 2-form, we derive the total mass and NUT charge of the Taub-NUT-AdS spacetime. Together with other thermodynamic quantities calculated through standard method, we conform the first law and Smarr relation. Then, we consider charged AdS NUTy spacetimes in Einstein-Maxwell theory with a cosmological constant, and show that the generalised Komar method works, too. We obtain all the thermodynamic quantities, and the first law and Smarr relation are checked to be satisfied automatically.

\vfill{\footnotesize jfliu@tju.edu.cn \quad hsliu.zju@gmail.com }


\thispagestyle{empty}
\pagebreak



\section{Introduction}
Taub-NUT spacetime is a simple solution of general relativity, which was constructed in the middle of last century \cite{nut1,nut2}. However, its thermodynamical properties remain unveiled due to its peculiar spacetime structure. Despite an event horizon, it contains two more Killing horizons at the south (north) pole associated with a string singularity, which is now called Misner string singularity\cite{misnerstring}.

The Misner string singularity can be rendered by imposing a periodic condition on the time coordinate and then the Misner string becomes unphysical\cite{misnerstring}. But, it was later shown that the time periodic condition is not necessary. The Taub-NUT spacetime is geodesically complete\cite{complete1,complete2}. In this paper, we shall not impose the periodic condition on time, but treat the time coordinate as real line, and thus the Misner strings remain physical.

Compared with Schwarzschild solution which has only one integration constant $m$, the Taub-NUT spacetime has one more integration constant $n$, which is usually called NUT parameter. The NUT charge related with parameter $n$ is usually viewed as a magnetic mass compared with the electric mass of black hole which is related with the mass parameter $m$ \cite{pd,mao}. But, the definition of NUT charge are still unclear, though there exist several trying of  defining NUT charge in order to get the differential form of the thermodynamic first law recently \cite{exa2,exa3,exa4,exa5,exa6,exa7,exa8,exa9}. In \cite{ww1,ww2,ww3,ww4,ww5,ptww}, more than one conserved charges associated with NUT parameter $n$ were introduced, whilst in \cite{nonut}, the first law of Kerr-Taub-NUT solution was established without introducing additional conserved charge associated with NUT parameter $n$.   Besides, there lacks in  defining the total mass of the NUTy spacetimes. Some treated the mass parameter $m$ as the total mass of the spacetime in the literature \cite{mass1,mass2,mass3,mass4}, it was noticed that the value of $m$ can be arbitrary negative and there exists no strong objection against the solution with negative $m$ to be physical \footnote{It is well known that the mass can be a little negative in the asymptotic AdS spacetime implied by Breitenlohner-Freedman(BF) bound \cite{bfbound}, but zero or the BF bound is just a number to the Taub-NUT-AdS case, the value of mass parameter $m$ can smoothly pass the point to a large negative number.   }, however, the total mass of the spacetime should be positive definite.

These two problems were addressed in \cite{llm} for Taub-NUT spacetime in Einstein gravity theories without cosmological constant. A systematic method of defining and calculating the NUT charge and total mass of Taub-NUT spacetime was proposed. Whilst in this paper, we want to check whether the method applies for Taub-NUT spacetimes in gravity theories with a cosmological constant.

In section 2, we consider Taub-NUT-AdS solutions in Einstein gravity plus a cosmological constant. We generalised the closed Komar form of pure Einstein gravity, and then find that the method works for Einstein gravity with a cosmological constant.  All the thermodynamic variables are derived and the first law is checked to be satisfied. In section 3, we go a step further to consider charged Taub-NUT-AdS spacetime in Einstein-Maxwell theory with a cosmological constant, and not surprisingly, we find the method works, too. We conclude our results in section 4.

\section{Taub-NUT-AdS}
We consider Taub-NUT solution in Einstein gravity theory with cosmological constant in four dimensional spacetime, the Lagrangian is simple
\be
{\cal L} = \sqrt g ( R - 2 \Lambda ) \,,
\ee
and the Taub-NUT-AdS solution is given by \cite{complete1,Hawking}
\be
ds_{\4}^2 = - f (dt + 2 n \cos \theta d\phi)^2 + \fft{dr^2}{f} + (r^2+n^2) (d\theta^2+\sin \theta^2 d\phi^2) \,,
\ee
with
\be
f=\fft{r^2- 2 m r - n^2}{r^2+n^2} - \fft{3 n^4 - 6 n^2 r^2-r^4}{l^2 (r^2+n^2)} \,.
\ee
Hereafter, we set $\Lambda = - \fft{3}{l^2}$. The solution has two integration constants, mass parameter $m $ and NUT parameter $n$, $l$ is a constant which is related to the cosmological constant $\Lambda$. When the NUT parameter $n$ vanishes, the solution goes back to Schwarzschild-AdS black hole.

The temperature and entropy can be obtained through the standard method
\be
T = \fft{1}{4 \pi r_+} \Big( 1 + \fft{3 (r_+^2 + n^2)}{l^2} \Big) \,, \qquad S = \pi (r_+^2 + n^2) \,.
\ee
$r_+$ is the radius of the event horizon, which is the largest root of $f(r)=0$. Additionally, there exist two other Killing horizons which locates in the south and north poles($\theta =0, \pi$). The corresponding two Killing vectors are given by
  \be
 l_\pm = \partial_\phi \mp 2 n \partial_t \,. \label{kvector}
  \ee
  It is worth pointing out that it has two advantages of written $l_\pm$ as above form instead of $\partial_t \pm 1/(2n ) \partial_\phi$. One is that (\ref{kvector}) has a smooth limit $n\rightarrow0$, the other is we can take $n$ as a thermodynamical potential compared with the form of Killing vector on the horizon of a Kerr black hole $\partial_t + \Omega \partial_\phi$, a detailed interpretation of this can be found in \cite{llm}.

  Though, many works have been done on the thermodynamics of Taub-NUT spacetime, the definitions of the total mass of the Taub-NUT-AdS spacetime and especially the NUT charge are not clear. The situation has changed due to recent work \cite{llm}, in which the mass of Taub-NUT spacetime and the NUT charge are defined and calculated through generalised Komar integration. Here, we shall apply the method to Taub-NUT-AdS spacetime. The key point of the method is that the Komar tensor is closed, $d * d \xi = 0$, where $\xi = \partial t$ is the Killing vector, in the Einstein gravity. But $d * d \xi$ doesn't vanish any more in the AdS case. A known result is
\be
-d*d\xi = - 2 * Ric \xi = - 2 \Lambda * \xi  \,,
\ee
the last step is obtained by using the equations of motion $R_{\mu\nu} = \Lambda g_{\mu\nu}$. We can see $d*d\xi$ is proportional to cosmological constant, and vanishes for pure Einstein gravity where $\Lambda = 0$.

So in the AdS case, one difficulty is to find a closed Komar-like form.
A direct calculation gives
\be
 -* d \xi = V(r) \Omega_\2 + U(r) dr \wedge ( dt + 2 n \cos \theta d \phi ) \,, \quad  \Omega_\2 = \sin \theta d\theta \wedge d\phi ,
\ee
with $U,V$ are
\be
V = (r^2 + n^2) f'  \,, \qquad U = \fft{2 n f}{r^2 + n^2} \,.
\ee
Thus
\be
-d *d\xi = ( V' + 2 n U ) \sin \theta dr \wedge d\theta \wedge d\phi = -2 \Lambda  \left(n^2+r^2\right) \sin \theta dr \wedge d\theta \wedge d\phi\,.
\ee
Again, we can see $-d*d\xi =0$ when cosmological constant vanishes.
Assuming there exist such a form $\omega$, whose Hodge dual takes the similar form as $*d\xi$
\be
 *\omega = \tilde V(r) \Omega_\2 + \tilde U(r) dr \wedge ( dt + 2 n \cos \theta d \phi ) \,,
\ee
that makes the combination of $-* d\xi$ and $*\omega$ is close
\be
d *(- d\xi + \omega ) = 0 \,.
\ee
And then we define this combination as the generalized Komar form
\be
Q = *(- d\xi + \omega ) \, ,
\ee
and  the closure of Q implies
\be
(\tilde V+V)' + 2 n (\tilde U+U) = \tilde V' + 2 n \tilde U - 2 \Lambda (r^2+ n^2) = 0 \,. \label{intcd}
\ee
It is worth pointing out that the $(\tilde U, \tilde V)$ can not be uniquely fixed through the above condition. It is easily seen that a 2-form $\lambda$, which satisfies $d*\lambda =0$,  can be added into $Q$, without changing the closure condition. As is shown in \cite{exa9,gauge}, a proper gauge choice is required to produce the consistent mass.

Following  the method in \cite{llm}, the NUT charge is defined by
\be
Q_N = \int_{r_+}^\infty (U+\tilde U) dr \,. \label{nutcharge}
\ee
In AdS case, the integration of $U$ is divergent, thus, our strategy for the choice of $\tilde U$ is that the divergent term of $U$ can be cancelled by $\tilde U$ and apart from this no additional term from $\tilde U$ emerges. One simple choice is
\be
\tilde U = (-2 n \fft{f(r)}{r})' \,,
\ee
and $\tilde V$ is thus given by
\be
\tilde V  = 4 n^2 \fft f r - \fft{6}{l^2} (\fft{r^3}{3} + n^2 r)  \,.
\ee
Now through (\ref{nutcharge}), the NUT charge can be calculated directly
\be
Q_N = \int_{r_+}^\infty (U+\tilde U) dr = \fft{n}{r_+} ( 1 + \fft{3 (n^2-r_+^2)}{l^2} ) \,.
\ee
The total mass of the Taub-NUT-AdS spacetime is then given by
\bea
M &=& \fft{\int d\phi}{8\pi}\Big( \int_{0}^{\pi} (V(r)+\tilde V(r))\sin \theta d\theta - \int_{r_+}^r 2 n \cos \theta (U(r')+\tilde U(r'))\Big|_{\theta=0}^{\theta=\pi} dr' \Big) \cr
&=& m + \fft {n^2}{r_+} ( 1 + \fft{3 (n^2-r_+^2)}{l^2} ) \,.
\eea
It is obvious that the mass and NUT charge has relation
\be
M = m + 2 \Phi_N Q_N \,. \label{mcrelation}
\ee
When letting $n \rightarrow 0$, the mass returns to the result of Schwarzschild mass, which gives a strong support for the correctness of our method. In the first looking, the mass can be negative for large $r_+$ due to the $-r_+^2$ term. But if we eliminate the $m$ by using $f(r_+) = 0$, we get
\be
M = \fft{3 n^4+r_+^4}{2 l^2 r_+}+\frac{n^2+r_+^2}{2 r_+}\,,
\ee
which is obviously non-negative.

In the AdS case, the form of the Killing vector on the south and north pole are not changed, thus the NUT potential remains the same as that of Einstein gravity
\be
\Phi_N = \fft n 2 \,.
\ee
At this stage, the first law can be checked and turns out to be satisfied
\be
\delta M = T \delta S + \Phi_N \delta Q_N \,.
\ee
The first law can also be generalised by taking the cosmological constant $\Lambda$ as a thermodynamic quantity $P = \fft{3}{8\pi l^2}$, and the corresponding thermodynamic volume is
\be
V = \fft{4 \pi}{3} r^3_+ ( 1 + \fft{3 n^2}{r_+^2} )\,.
\ee
Interestingly, the thermodynamical volume is the same as that in \cite{free}.

The first law is then given by
\be
\delta M = T \delta S + \Phi_N \delta Q_N + V\delta P\,.
\ee

As shown in \cite{llm}, the Smarr relation can be obtained by using $dQ=0$
\bea
\fft{1}{8 \pi} \int_\Sigma Q[\xi] &=& \fft 14 \big[ ( V+ \tilde V )|_{r_+}^\infty + \int_{r_+}^\infty 2 n (U(r')+\tilde U(r')) dr' \big]  \cr
 &=& \fft 14 \big[  V |_{r_+}^\infty + \int_{r_+}^\infty (2 n U(r')+2 \Lambda (r'^2+n^2)) dr' \big]= 0 \,.
\eea
The second line is obtained by using the integrability condition (\ref{intcd}), and it is worth pointing out that it is not necessary to make a definite choice for $(\tilde U, \tilde V)$ to  get the Smarr formula from $dQ$ = 0. Explicitly, in the infinity we get
\be
 V+ \int 2 n (U(r) +2 \Lambda (r^2+n^2))dr) |_{r\rightarrow \infty} = 2 m \,,
 \ee
 whilst, on the horizon
 \be
( V+ \int 2 n (U(r') +2 \Lambda (r'^2+n^2))dr')|_{r_+}  = 4 T S - 4 P V - 4 \Phi_N Q_N \,.
 \ee
 Then putting them together, we get the exact form of Smarr relation of Taub-NUT-AdS spacetime
 \be
 M = 2 (T S - P V)\,,
 \ee
the NUT charge and potential doesn't contribute directly to the Smarr formula as the same as that in Einstein gravity.
The Gibbs free energy is
\be
F = M - T S - \Phi_N Q_N = \fft m 2 - \fft{1}{2 l^2} \big( 3 n^2 r_+ + r_+^3 \big)\,,
\ee
which is consistent with Euclidean action\cite{free}.

So far, we consider the Taub-NUT solution with symmetrically distributed  Misner strings. Our method can easily be generalised to the asymmetric case. The solution with asymmetric Misner strings can be obtained through a linear coordinate transformation
\be
t\rightarrow t- 2 n \alpha \phi \,, \quad \phi \rightarrow \phi \,,
\ee
where the parameter $\alpha$ is a real constant. Then the Killing vectors at the north and south pole change to
\be
l_\pm = \partial_\phi \mp 4 \Phi_N^\pm \partial_t \,, \quad \Phi_n^\pm = \fft 12 n (1\pm \alpha)\,.
\ee
And thus the distribution of Misner strings affect the NUT potential, too. When $\alpha = 0$, the Misner strings are symmetric and the north and south poles are in the equal foot. When $\alpha = 1 $, the Misner sting only emerges at the south pole, whilst $\alpha = -1$ disappears at the south pole.

Though, the NUT potential has a direct contribution from $\alpha$, the rest thermodynamical quantities, mass, NUT charge, temperature and entropy are not modified by $\alpha$. The first law becomes
\be
\delta M = T \delta S + \Phi_N^+ \delta Q_N^+ + \Phi_N^- \delta Q_N^-\,,
\ee
where the NUT charge at south and north poles are the same
\be
Q_N^\pm  = \fft{n}{2 r_+} ( 1 + \fft{3 (n^2-r_+^2)}{l^2} ) \,.
\ee
Since $Q_N^\pm$ are the same, the parameter $\alpha$ in the last two terms  $\Phi_N^+ \delta Q_N^+ + \Phi_N^- \delta Q_N^- $ of the first law will cancel out, and $\alpha $ will not appear in the first law. Note that there is a factor $2$ in the denominator of NUT charges. When $\alpha = 0$, the south and north poles have the same NUT potential $\Phi_N^\pm  = \Phi_N $, thus the south and north pole NUT charges can be recognized as the same class, and they can be summed as a whole NUT charge $Q_N^++Q_N^- = Q_N$, then we recover results of the symmetric distributed Misner strings.

\section{Dyonic Taub-NUT-AdS}
In this section we generalise the method to charged case in theory of  Einstein-Maxwell gravity plus a cosmological constant
\be
L = \sqrt g ( R -2 \Lambda- F^2) \,,
\ee
where $F=dA$.
The solution is
\be
ds_\4^2 = -f(dt + 2 n \cos \theta d\phi)^2 + \fft{dr^2}{f} + (r^2+n^2) (d\theta^2+\sin \theta d \phi^2) \,,
\ee
with
\be
f = \fft{r^2- 2 m r - n^2+e^2+g^2}{r^2+n^2} - \fft{3 n^4 - 6 n^2 r^2-r^4}{l^2 (r^2+n^2)} \,,
\ee
and the Maxwell field is
\be
A =  - g \cos \theta d\phi + \fft{(g n + e r )(dt + 2 n \cos \theta d \phi)}{r^2+n^2} \,.
 \ee
The corresponding dual filed is $\tilde F = d B = * F$ with gauge potential $B$  given by
\be
B = e \cos \theta d\phi - \fft{( e n - g r ) (dt + 2 n \cos \theta d\phi)}{r^2+n^2} \,.
\ee
Beyond the mass and NUT parameters $(m,n)$, there exist two additional integration parameters $(e,g)$ which correspond to the electric and magnetic charges. The spacetime has an event horizon at $r=r_+$, with $r_+$ as the largest root of $f(r)=0$. There exist two other Killing horizons, too, the associated Killing vectors are the same as the neutral case (\ref{kvector}). Thus, we have the same NUT potential
\be
\Phi_N=\fft n 2 \,.
\ee

The temperature and entropy of the charged NUTy spacetime can be obtained through the standard method as before,  and they are given by
\be
T = \fft{1}{4 \pi r_+} ( 1 + \fft{3 (r_+^2 + n^2)}{l^2} - \fft{e^2+g^2}{r_+^2 + n^2} ) \,, \quad S = \pi (r_+^2 + n^2) \,.
\ee

Turn to the thermodynamics of the charged NUTy spacetime, we first need to obtain the closed 2-form $Q$. Without cosmological constant, the combination in Einstein-Maxwell theory,
\be
-*d\xi - *F A_\lambda A^\lambda - * \tilde F B_\lambda B^\lambda
\ee
is closed\cite{llm}, where $\xi$ is a Killing vector. Again, due to the inclusion of the cosmological constant, the combination is not closed anymore, and thus an additional term is required as before. Fortunately, the recipe is the same as the case in the previous section, too. We find that
\be
Q =  -*d\xi  -  *F A_\lambda A^\lambda -  * \tilde F B_\lambda B^\lambda +  * \omega
\ee
is closed, with
\be
* \omega = \tilde V(r) \Omega_\2 + \tilde U(r) dr \wedge ( dt + 2 n \cos \theta d \phi ) \,,
\ee
and $\tilde U$,$\tilde V$ are under constraint
\be
\tilde V' + 2 n \tilde U - 2 \Lambda (r^2+ n^2) = 0 \,.
\ee
It is worth mentioning that the above expression is derived by using equation of motions and the constraint $(\tilde U, \tilde V)$ is the same as that of the previous section. We make the same choice of $(\tilde U , \tilde V)$
\be
\tilde U = (\fft{-2 n f}{r})' \,, \qquad \tilde V = \fft{4 n^2 f}{r} + \fft 23 \Lambda r (r^2+3 n^2) \,,
\ee
the closure of $Q$, $dQ=0$, is satisfied, too. Though we don't uniquely derive the expression of $(\tilde U , \tilde V)$, but choose a simple form, it turns out to be general, at least it works for both neutral and charged cases.

With the generalised Komar 2-form $Q$, we can now define the total mass and NUT charge through the same procedure and obtain
\bea
M &=& \fft{\int d\phi}{8\pi}\Big( \int_{0}^{\pi} (V(r,\theta)+\tilde V(r,\theta)) d\theta - \int_{r_+}^r 2 n \cos \theta (U(r',\theta)+\tilde U(r',\theta))\Big|_{\theta=0}^{\theta = \pi} dr' \Big) \cr
&=& m + \fft {n^2}{r_+} ( 1 - \fft{3 (r_+^2-n^2)}{l^2}  - \fft{e^2+g^2}{r_+^2 + n^2}) \,,\cr
Q_N&=&\fft {n}{r_+} \big( 1 - \fft{3 (r^2-n^2)}{l^2}  - \fft{e^2+g^2}{r_+^2 + n^2} \big) \,.
\eea
It can be easily seen that the mass and NUT charge relation (\ref{mcrelation}) holds, too.

With the equation of motion of Maxwell field $d*F = 0 $ and the Bianchi identity $dF=0$, we can calculate the electric and magnetic charges through the same strategy
\bea
Q_e &=& - \fft12\big( \int^\pi_{0} \tilde F_{\theta \phi}(r,\theta') d\theta' - \int^r_{r_+} \tilde F_{r\phi} (r',\theta)\big|^{\theta=\pi}_{\theta=0} dr' \big) = - \fft 12 B_\phi(r_+) \big|_{\theta=0}^{\theta = \pi} \cr
&=& e + 2 n \fft{gr_+ - e n}{r_+^2+n^2} \,, \cr
Q_g &=&  \fft12( \int^\pi_{0}  F_{\theta \phi}(r,\theta') d\theta' - \int^r_{r_+}  F_{r\phi} (r',\theta)\big|^{\theta=\pi}_{\theta=0} dr' ) =  \fft 12 A_\phi(r_+) \big|_{\theta=0}^{\theta = \pi}\cr
&=& g - 2 n \fft{e r_+ + g n }{r_+^2+n^2}\,.
\eea
The electric and magnetic potential are defined by
\be
\Phi_e = \xi^\mu A_\mu\big|^{r_+}_\infty =  \fft{e r_+ + g n}{r_+^2+n^2} \,, \quad \Phi_g = \xi^\mu B_\mu\big|^{r_+}_\infty = \fft{g r_+ - e n}{r_+^2+n^2} \,.
\ee

Similar to NUT charge, we can also define NUT induced electric and magnetic charges through $d*F=0$ and $dF=0$ at the south and north poles
\bea
Q_{eN} &=& \int_{r_+}^\infty \tilde F_{tr} dr = \fft 12 B_t \big|_{\infty}^{r_+}  = \frac{  g r_+ - e n }{n^2+r_+^2} \,,  \cr
Q_{gN} &=& \int_{r_+}^\infty  F_{tr} dr = \fft 12 A_t(r_+) \big|_{\infty}^{r_+} =  \frac{e r_+ + g n}{n^2+r_+^2} \,,
\eea
and the corresponding potential are defined through the analogous method of defining the electric and magnetic potentials
\bea
\Phi_{eN} &=& \fft 14 l^\mu(A_\mu(\theta=0)+A_\mu(\theta=\pi))\big|^{r_+}_\infty = - \frac{n(e r_+ + g n)}{n^2+r_+^2} \,,  \cr
\Phi_{gN} &=& - \fft 14 l^\mu(B_\mu(\theta=0)+B_\mu(\theta=\pi))\big|^{r_+}_\infty = \frac{n(g r_+ -e n)}{n^2+r_+^2} \,.
\eea

At this stage, we derived all the thermodynamical quantities and they are summarized as follow.
\bea
&&T = \fft{1}{4 \pi r_+} ( 1 + \fft{3 (r_+^2 + n^2)}{l^2} - \fft{e^2+g^2}{r_+^2 + n^2} ) \,, \quad S = \pi (r_+^2 + n^2) \,,  \cr
&& M = m + \fft {n^2}{r_+} ( 1 - \fft{3 (r_+^2-n^2)}{l^2}  - \fft{e^2+g^2}{r_+^2 + n^2}) \,, \cr
&& Q_N =  \fft {n}{r_+} ( 1 - \fft{3 (r_+^2-n^2)}{l^2}  - \fft{e^2+g^2}{r_+^2 + n^2} ) \,,\qquad \Phi_N = \fft n 2 \,,  \cr
&& Q_{eN} = \fft{g r_+ - e n}{r_+^2 + n^2} \,, \qquad  \Phi_{eN} = - \fft{n(e r_+ + g n)}{r_+^2 + n^2} \,, \cr
&& Q_{gN} = \fft{e r_+ + g n}{r_+^2 + n^2} \,, \qquad \Phi_{gN} = \fft{n(g r_+ - e n)}{r_+^2 + n^2} \,, \cr
&& Q_{e} = e + 2 n Q_{eN} \,, \qquad  \Phi_{e} =  \fft{(e r_+ + g n)}{r_+^2 + n^2} \,, \cr
&& Q_{g} = g - 2 n Q_{gN}\,, \qquad \Phi_{g} = \fft{(g r_+ - e n)}{r_+^2 + n^2} \,.
\eea
When setting $ l\rightarrow \infty $, these quantities turn back to the result of Einstein-Maxwell case without cosmological constant as expected.

And it can be checked that the first law is automatically satisfied
\be
\delta M = T \delta S   + \Phi_N \delta Q_N + \Phi_e \delta Q_e+ \Phi_g \delta Q_g + \Phi_{eN} \delta Q_{eN}+ \Phi_{gN} \delta Q_{gN}\,.
\ee
Again, when we treat the cosmological constant as a thermodynamical variable, the corresponding thermodynamical volume can also be derived,
\be
P = \fft{3}{8 \pi l^2} \,, \qquad V  = \fft{4 \pi}{3} r^3_+ ( 1 + \fft{3 n^2}{r^2} )   \,.
\ee
And the generalised first law is
\be
\delta M = T \delta S   + V \delta P + \Phi_N \delta Q_N + \Phi_e \delta Q_e+ \Phi_g \delta Q_g + \Phi_{eN} \delta Q_{eN}+ \Phi_{gN} \delta Q_{gN}\,.
\ee
From $dQ=0$, we can derive the Smarr relation
\be \label{smarr}
M = 2 ( T S - P V) + \Phi_e Q_e + \Phi_g Q_g\,.
\ee

The Free energy can be evaluated through the Euclidean action $G = I/\beta$, with $\beta $  is the inverse of the temperature, $\beta = 1/T$. The full action is given by
\be
I = \fft{1}{16 \pi} \int_M d^4x \sqrt g \big( R + \fft{6}{l^2} - F^2 \big) + \fft{1}{8 \pi} \int_{\partial M} d^3 x \sqrt h \big( {\cal K} - \fft 2 l - \fft l 2 {\cal R}(h) \big)\,,
\ee
where, $h$ is determinant of the induced metric, ${\cal K}$ is the extrinsic curvature and ${\cal R} (h)$ is the boundary Ricci scalar. Substituting the solution into the whole action, we can obtain the free energy
\be
G = \fft{m}{2} - \fft{r_+\big( 3 n^2 + r_+^2 )}{2 l^2} - \fft{r_+( (e^2-g^2)(r_+^2 - n^2) + 4 e g n r_+  \big)}{2 (r_+^2 + n^2)^2} \,.
\ee
This result is consistent with our newly derived thermodynamical quantities
\be
F = M - T S - \Phi_N Q_N - \Phi_e Q_e - \Phi_{eN} Q_{eN} \,.
\ee

Finally, it can be easily checked that the first law has electromagnetic duality under
\be
e \rightarrow g \,, \qquad g \rightarrow - e \,.
\ee
As in the previous section, the method can be easily generalized to asymmetric distributed Misner stings, and the results are similar.

\section{Conclusions}
 Though the Taub-NUT-AdS solution exists for years, the thermodynamical properties has not been totally understood. Many works has been done on the first law of Taub-NUT spacetimes, but there still lack in uniquely deriving the NUT charge. Recently, a systematic way of defining and calculating the NUT charge and the total mass of the Taub-NUT spacetime has been proposed. In this paper, we apply this method to Taub-NUT-AdS spacetime in Einstein gravity plus cosmological constants. A key ingredient of this method is to get a closed 2-form $Q$. For, pure Einstein gravity, the closed 2-form is just the derivative of Killing vector $-*d\xi$, it is no longer closed for Einstein gravity plus cosmological constant. To solve this problem, we construct a generalised closed 2-form by introducing an additional term. Then starting with this generalised 2-form, we obtain the NUT charge and total mass of Taub-NUT-AdS spacetime. Together with the entropy and temperature which can be derived through standard method, the first law is checked to be satisfied. The Smarr relation can be obtained though the closure of the generalised 2-form as usual.

Then, we turn to the Einstein-Maxwell theory with a cosmological constant. The same problem, that the usual closed 2-form in Einstein-Maxwell without cosmological constant is no longer closed, emerges. Fortunately, the recipe for this problem is similar, too. Especially, the expression of the additional term needed to construct the new generalised 2-form has the same expression, when written in terms of metric functions. With this generalised closed 2-form, we derive the NUT charge and total mass of the spacetime. Since Maxwell's equation of motion and Bianchi identity, $*F_\2$ and $F_\2$ are closed, too. We can calculate electric and magnetic charges through the 2-forms, analogous to the NUT charge, the NUT-induced charges can also be derived. Finally, we present all the thermodynamic quantities and  the first law and Smarr relation of the spacetime are indeed satisfied. We calculate the Free energy of the system by using the derived thermodynamic variables and the results are consistent with that of Euclidean action.

We mainly studied the thermodynamics of the Taub-NUT-AdS spacetimes, it will have fruitful applications in holography, such as holographic complexity, conductivity, viscosity eta. Especially, the effect of NUT charge, together with NUT induced electric and magnetic charges, in holography is worth exploring.

\section*{Acknowledgement}

We are grateful to Hong Lu and Liang Ma for useful discussions. This work is supported in part by NSFC (National Natural Science Foundation of China) Grants No.~12075166.

\end{document}